\newif\ifcqg\cqgtrue
\ifcqg
\documentclass[]{iopart}
\usepackage{iopams}
\usepackage{setstack}
\eqnobysec

\else
\documentclass[aps,letter,prd,thightenlines,nofootinbib,showpacs,eqsecnum,amsfonts,amssymb,amsmath]{revtex4}
\usepackage{dcolumn}
\usepackage{bm}           % bold math

\fi

\ifcqg

\else

\fi

\usepackage{epsfig}
\usepackage{dcolumn}% Align table columns on decimal point
\usepackage{bm}% bold math
\DeclareGraphicsExtensions{.eps,.ps,.eps.gz,.ps.gz,.eps.Z,{}}

\newcommand{\mg}{\langle}
\newcommand{\md}{\rangle}

\newcommand{\sym}{\hbox{sym.}}

\newcommand{\ad}{a^{\dagger}}
\newcommand{\nue}{\nu_{\rm eff.}}

\newcommand{\Mpl}{M_{\hbox{\tiny{Pl.}}}}

\newcommand{\vk}{{\bf k}}

\newcommand{\vx}{{\bf x}}

\newcommand{\dd}{{\rm d}}
\newcommand{\ii}{{\rm i}}
\newcommand{\mG}{{\cal G}}

\newcommand{\mO}{{\cal O}}
\newcommand{\mP}{{\cal P}}
\newcommand{\mQ}{{\cal Q}}

\newcommand{\mU}{{\cal U}}
\newcommand{\mT}{{\cal T}}

\newcommand{\Id}{{\rm Id}}
\newcommand{\lvac}{\langle0\vert}
\newcommand{\rvac}{\vert0\rangle}

\newcommand{\hu}{{\bf \hat{u}}}
\newcommand{\hs}{{\bf \hat{s}}}
\newcommand{\hd}{{\bf \hat{d}}}

\newcommand{\cs}{\hbox{\textbf{s}}}
\newcommand{\cd}{\hbox{\textbf{d}}}
\newcommand{\cu}{\hbox{\textbf{u}}}
\newcommand{\erfi}{\hbox{Erfi}}
\newcommand{\textEi}{{\rm Ei}}

\newcommand{\mOd}{{\cal Q}}
\newcommand{\Dirac}{\delta_{\rm D}}

\begin{document}
\title{Mode coupling evolution in arbitrary inflationary backgrounds}
%\title{Superhorizon correlation properties of test fields in arbitrary inflationary backgrounds}

\author{Francis Bernardeau}
\ifcqg
\ead{francis.bernardeau@cea.fr}
\address{CEA, Institut de Physique Th{\'e}orique, 91191 Gif-sur-Yvette c{\'e}dex, France 
CNRS, URA-2306, 91191 Gif-sur-Yvette c{\'e}dex, France}
\else

\author{Francis Bernardeau}
\email{francis.bernardeau@cea.fr}
\affiliation{CEA, Institut de Physique Th{\'e}orique CEA/DSM/IPhT,\\
         CNRS, URA-2306, 91191 Gif-sur-Yvette c{\'e}dex, France.}
\fi


\date{\today}

\begin{abstract}
The evolution of high order correlation functions of a test scalar field in arbitrary inflationary backgrounds is computed. Whenever possible, exact results are derived from quantum field theory calculations. 
Taking advantage of the fact that such calculations can be mapped, for super-horizon scales, into those of a classical system, we express the expected correlation functions in terms of classical quantities, power spectra, Green functions, that can be easily computed in the long-wavelength limit. Explicit results are presented that extend those already known for a de Sitter background. In particular the expressions of the late time amplitude of bispectrum and trispectrum, as well as the whole high-order correlation structure, are given in terms of the expansion factor behavior. 

When compared to the case of a de Sitter background, power law inflation and chaotic inflation induced by a massive field are found to induce high order correlation functions the amplitudes of which are amplified by almost one order of magnitude. These results indicate that the dependence of the related non-Gaussian parameters -- such as $f_{\rm NL}$ -- on the wave-modes is at percent level.  
\end{abstract}

\ifcqg
% \maketitle
\else
\maketitle
\fi

\section{Introduction}

It is now clearly established that standard single field inflation cannot produce
significant non-Gaussianities (NG) during or immediately after the inflationary phase. This has been explicitly shown by 
Maldacena in Ref.  \cite{2003JHEP...05..013M} where it clearly appears that standard single
field inflation leads to no or very little primordial non-Gaussianities.

Multiple-field inflation has now long been recognized as a possible mechanism for the generation of primordial metric
NG fluctuations (see recent review papers in Refs \cite{2010CQGra..27l4001K,2010CQGra..27l4007L,2010CQGra..27l4002W,2010CQGra..27l4004B}). The exploration of the various types of coupling terms that appear in the action  then leads to distinguish 
gravity from non-gravity mediated couplings (see \cite{2010CQGra..27l4004B} for recent review on the origin of this distinction). This decomposition comes from the various behaviors of the terms that are present in the third order action.
In one-field inflation, because the field fluctuations and the metric fluctuation are locked together, only the former can be found. In multiple-field inflation however this is not necessarily so. In particular isocurvature degrees of freedom open the possibility 
of having a richer phenomenology. And whereas gravity mediated couplings are ubiquitous but induce only modest effects \cite{2005JCAP...09..011S,2006JCAP...05..019V}, non-gravity mediated couplings can be very efficient, although nothing ensures that they are generically at play (this is at the heart of Refs. \cite{2007PhRvD..76d3526B,2003PhRvD..67l1301B,2002PhRvD..66j3506B}) .

To make such perturbations play a role, one indeed needs a mechanism to transfer isocurvature modes into adiabatic
fluctuations. For instance the curvaton model is based on the survival of (massive) isocurvature modes until late after the end of inflation that can alter the subsequent expansion history of the universe \cite{2002PhLB..524....5L}. This is a particular case of modulated inflation \cite{2003astro.ph..3614K,PhysRevD.69.023505,2004PhRvD..70h3004B}. Other mechanisms assume that isocurvature modes can change the end-point of inflation or alter the (p)-reheating sequences (as in \cite{2009PhRvL.103g1301B}). Such mechanisms can also happen in the context of hybrid inflation. This latter situation is in particular advocated in Refs. \cite{1990PhRvD..42.3936S,2002PhRvD..65j3505B,2002PhRvD..66j3506B,2003PhRvD..67l1301B} where isocurvature modes are shown to be able to induce large NGs in the metric fluctuations. 

In this paper we assume that such a mechanism is at play and focus our analysis on the intrinsic statistical properties of 
isocurvature modes, that is of modes that do not participate in the metric fluctuations at the time we are interested in. 
Even though this is a simple setting, exact results are still difficult to obtain since they depend on the actual dynamics of the expansion of the universe. Some exact results have been obtained in case of a de Sitter background (in Ref. \cite{1993ApJ...403L...1F} bispectrum of a field with a cubic potential
 is derived; in Ref. \cite{2003JHEP...05..013M} more complicated types of cubic interaction are considered and in Ref. \cite{2004PhRvD..69f3520B} the four-point function is derived for quartic interactions) that has been the favorite playground for such calculations.  The evolved bispectra,  trispectra, have in particular been found to grow like the number of efolds after horizon crossing. This is an important result that lays the foundation
for the computation of the mode dependence of the coupling parameters (which are usually expressed in terms of $f_{\rm NL}$ parameter
as introduced in \cite{2002astro.ph..6039K}). This possibility is now being explored (as in Ref. \cite{2009arXiv0911.2780B} for instance for some classes of model.)
However, though inflationary models lead to background evolution that are generically close to a de Sitter, this is only an approximation, and which at best can be valid for only a limited period of time.

In this article we aim at being as much exhaustive as possible in presenting exact results on the evolution of the statistical properties of a test scalar field, with non-vanishing self interaction terms, in arbitrary inflationary background. We will make use of the relation between quantum and classical evolutions. However we will not use  explicitly  the $\delta N$ formalism  (as introduced in \cite{1982PhLB..117..175S,1996PThPh..95...71S} and see Ref. \cite{2005PhRvL..95l1302L} in the context of nonlinear expansions) as it does not necessarily give controlled approximations but try to be as much precise as possible to draw the line between results that are of sub-Hubble origin and those that can be accounted for in a super-Hubble classical calculation. In the ``tree-theorem'' in Ref. \cite{2008PhRvD..78f3534W}  Weinberg demonstrated that there exists a classical system that exhibits the same
tree order correlations properties as the quantum system. Although we will not exactly use this solution, it will eventually lead us to a complete description of the super-Hubble tree order correlation functions of the field.

The paper is divided as follows. In section 2 we recall the method and the known results in case of a de Sitter background and describe in some details the late time super-Hubble limit of bispectra and trispectra. In the following section we make explicit the connection between calculations of correlation functions in a quantum context and classical calculations of stochastic field evolution. It shows in particular that the late time evolution of correlation functions can be computed in a classical context with the introduction of the Green function of the classical
evolution of the free fields. This observation is exploited to obtain new results for the computation of the late time behavior of correlation functions in arbitrarily backgrounds. These results are presented in the section 4. In section 5 the resulting correlation functions are presented in a systematic way with the description of the tree structure of the field correlation functions at leading order in perturbation calculations.

\section{Mode coupling computation}

\subsection{General framework}

We are interested here in the evolution equations governing a test scalar field $\chi(\vx,\eta)$ in the homogeneous background
of an inflationary universe. Throughout the paper we assume that the spatial curvature vanishes so that the background in which the field evolves can be described with the following metric,
\begin{equation}
\dd s^2=a^2(\eta)(-\dd \eta^2+\dd x^2)
\end{equation}
and $a(\eta)$ determines the background expansion. The time dependence of $a(\eta)$ is assumed to be given by a sector of the 
theory which is independent of $\chi$. We neglect in particular here gravity induced loop effects such as the backreaction of the inhomogeneities in $\chi$ on the background evolution. The $\chi$ field therefore behaves as a minimally coupled test field. We further assume that the interaction operator of $\chi$, $V\left[\chi\right]$, is a function of $\chi$ only. This is again for convenience since the method we introduce could be a priori extended to more more complicated operators\footnote{As this is the case for the cubic part of the action for the inflaton field for instance.}. 
The aim of the following calculation is then to explore the statistical properties of the $\chi$ field or more precisely of its
Fourier modes $\chi_{\vk}(\eta)$, 
\begin{equation}
\chi_{\vk}(\eta)=\int{\dd^3\vx}\,\exp(-\ii\vk.\vx)\,\chi(\vx,\eta),
\end{equation}
that is compute the late time power spectrum of the latter, $\mP_{\chi}(k)$, defined as
\begin{equation}
\langle \chi_{\vk}\chi_{\vk'}\rangle=(2\pi)^3\Dirac(\vk+\vk')\,P_{\chi}(k)
\end{equation}
where $\Dirac$ is the 3-dimensional Dirac delta function
and its higher order correlation functions such as the bispectrum, $B_{\chi}(\vk_{1},\vk_{2},\vk_{3})$, defined as
\begin{equation}
\langle \chi_{\vk_{1}}\chi_{\vk_{2}}\chi_{\vk_{3}}\rangle=(2\pi)^3\Dirac(\vk_{1}+\vk_{2}+\vk_{3})\,B_{\chi}(\vk_{1},\vk_{2},\vk_{3}).
\end{equation}
For convenience we also define the reduced high order correlation functions such as $Q_{3}(\left\{k_{i}\right\})$ or 
$Q_{4}(\left\{k_{i}\right\})$ generally as
\begin{equation}
\mg \chi_{\vk_1}\ldots\chi_{\vk_p}\md_c=(2\pi)^3\Dirac\left(\sum_i\vk_i\right)\,Q_{p}(\left\{k_{i}\right\})\sum_{i} \prod_{j \ne i} P_{\chi}(k_{j})\label{Qpdef}
\end{equation}
and our final results will be expressed in terms of $Q_{3}$ or $Q_{4}$ functions for simple potentials.

For the actual resolution of the equations, it is convenient to introduce the reduced field $u(\vx,\eta)$ as
\begin{equation}
u(\vx,\eta)=a(t)\chi(\vx,\eta).
\end{equation}
The  resulting motion equation is then,
\begin{equation}
u''(\vx,\eta)-\frac{a''}{a}u(\vx,\eta)-\Delta\,u(\vx\eta)=-a^3\,\frac{\dd V}{\dd \chi}[u(\vx,\eta)/a]\label{uevol1}
\end{equation}
The right-hand side of this equation contains nonlinear terms in $u(\vx,\eta)$ and this evolution equation cannot then be solved 
explicitly. One can then define the free field $\hu^{(0)}$ which is the field that satisfies the motion equation when only the quadratic terms in the Hamiltonian are kept. 

In the following we put an $\hat{\ }$ over $\cu$, or on other fields, when we insist on the fact that it can be viewed as quantum operators.
It is then assumed that the field $\hu^{(0)}$ can be decomposed into creation and annihilation operators, $a_{\vk}$ and $\ad_{\vk}$ such that
\begin{equation}
\hu^{(0)}(\vx,\eta)=\int\frac{\dd^3\vk}{(2\pi)^3}\,\exp(\ii\vk.\vx)\,\hu^{(0)}_{\vk}(\eta)
\end{equation}
with
\begin{equation} \hu^{(0)}_{\vk}(\eta)=
u^{(0)}_{k}(\eta)\,a_{\vk}+{u^{(0)}_{k}}^{*}(\eta)\,\ad_{-\vk}\label{u0exp1}
\end{equation}
where we have
\begin{equation}
[a_{\vk},\ad_{\vk'}]=\hbar\,(2\pi)^3\Dirac(\vk-\vk').
\end{equation}
In Perturbation Theory one makes the assumption that the field $\hu(\vx,\eta)$ can be expanded in terms of those same operators. The time dependence of the function $u^{(0)}_{k}(\eta)$ is set by the time dependence of the background. In the following we will be interested in cases where the mass of the free field vanishes (or is assumed to be small enough compared to the Hubble constant) so that the motion equation is,
\begin{equation}
{u^{(0)}_{k}}''(\eta)-\frac{a''}{a}u^{(0)}_{k}(\eta)+k^2\,u^{(0)}_{k}(\eta)=0.\label{u0evol}
\end{equation}
In principle this equation can be solved once $a(\eta)$ is known. 

In the following, we will then assume a simple form for the potential. 
\begin{equation}
V(\chi)=\frac{\lambda}{p!}\ \chi^p\label{Vshape}
\end{equation}
where $p$ is an integer, $p\ge 3$, which in practice will be either 3 or 4 corresponding to respectively a cubic or a quartic coupling.

\subsection{Conditions for large isocurvature fluctuations}
\label{Conditionsonlambda}

It is probably worth investigating in some details the conditions (on $\lambda$) for $\chi$ to develop
fluctuations in the first place. 
A priori such a potential is bound to induce an effective mass square to the field $\chi$ of about ${\lambda}\,H^{p-2}$
so that one should have $\lambda\lesssim H^{4-p}$ in order to fluctuations in the $\chi$-field to develop. For a cubic coupling that implies that $\lambda\lesssim H$ and $\lambda\lesssim 1$ for a quartic coupling. The derivation of this constraint
can be made more quantitative with the help of
a classical stochastic approach (introduced in \cite{1994PhRvD..50.6357S}) which can be used to infer the
probability distribution function of the value of $\chi$ at horizon crossing. 
%%%%%%%%%%%%%%%%%%%%%%%%%%%%%%%%%%%%%%%%
The Fokker equation
for the time dependent probability distribution function of $\chi$, $\mP(t,\chi)$ can be
derived from the evolution equation of $\chi$. It is given by,
\begin{eqnarray}
\frac{\partial \mP}{\partial t}&=&\frac{H^3}{8\pi^2}
\frac{\partial^2 \mP}{\partial \chi^2}
+\frac{1}{3H}\frac{\partial}{\partial \chi}
\left(\frac{\partial V(\chi)}{\partial \chi}\mP\right),
\end{eqnarray}
where $t$ is the physical time. This equation has a non trivial late time, and time independent, solution
given by (for $p=4$),
\begin{equation}
\mP(\chi)=\frac{1}{2\Gamma(5/4)H}\left(\frac{\pi^2\lambda}{9}\right)^{1/4}
\exp\left[-\frac{\pi^2\lambda\,\chi^4}{9 H^4}\right].
\end{equation}
It shows that the typical excursion values for $\chi$ are of the order of $\lambda^{-1/4}H$ for a quartic coupling.
It corresponds to excursion values of a field of mass of the order of $\lambda^{1/4}H$.
Modes of a given wavelength, whose amplitude if the field is effectively massless, are therefore not suppressed  if 
$\lambda\lesssim 1$. 

This result can be further comprehended from the expression of 
loop corrections to the two-point correlation of self coupled a test field. In \cite{2005PhRvD..71f3529B}
it is shown in particular that a complex scalar field with a quartic coupling imbedded in a supersymmetric multiplet
developed an effective mass square of the order of $\lambda H^2$ times a logarithmic divergent term (which in the context 
of inflation is reasonably finite.)
This is, if one needs to be fully specific,  the context that we have in mind in the investigations we present here.
Note that this condition ensures that the isocurvature fluctuations are of the order of $H$ at the time of horizon crossing.
It does not prejudge the conditions for the survival of the isocurvature perturbations till the end of inflation (where it is assumed that
some sort of mode transfer can take place.). For de Sitter background perturbation theory results suggest that
the condition $\lambda\lesssim H^{4-p}$ should be replaced by $\lambda N_{e}\lesssim H^{4-p}$ where $N_{e}$
is the number of efolds between horizon crossing and the end of inflation. We will see at the very end of the paper
how this condition is to be (slightly) changed for other backgrounds.

\subsection{The ``in-in'' formalism}

One unambiguous way to derive the statistical properties of the field $\chi$ is to use the ``in-in'' formalism from Schwinger and Keldysh (see Refs. \cite{1961PNAS...47.1075S, Keldysh:1964ud} and \cite{2005PhRvD..72d3514W} for a presentation and use in a cosmological context). We briefly present its implementation in this context here. 
%Let us consider an observable $\mQ(\eta)$, that is any function of the field $\chi(\vx,\eta)$. Typically we will consider cases where $\mQ$ is an equal time product of fields. What we want to do is to exhibit an operator that allows to express $\mQ(t)$ as a function of $\mQ^{(0)}(\eta)$, which is the same formal observable but computed with the free fields values.

The first step is to decompose the Hamilonian of the system in a classical part, the one responsible for the background evolution, and a quantum part that will govern the evolution of the quantum fluctuations. The latter contribution can be further split in two parts,\begin{equation}
H_{Q}=H_{Q}^{(0)}+H^{(I)},
\end{equation} 
one which is quadratic in the fields, $H_{Q}^{(0)}$, and which governs the evolution of the free fields, and an interaction term, $H^{(I)}, $which is supposed to be small so that the actual evolution of the fields remain close to the free field evolution. This is obviously a basic hypothesis for doing perturbation calculations.
This decomposition allows to define the operator, $\mU^{(I)}(\eta,\eta_0)$,  as
\begin{equation}
\frac{\dd }{\dd \eta}\mU^{(I)}(\eta,\eta_0)=-\ii H^{(I)}[\eta,\chi^{(0)}(\eta)]\,\mU^{(I)}(\eta,\eta_0)\label{mUEvol}
\end{equation}
where the interaction Hamiltonian is written in terms of the free fields. The solution of (\ref{mUEvol}) formally reads,\begin{equation}
\mU^{(I)}(\eta,\eta_0)=\mT\left\{
\exp\left(-\ii \int_{\eta_{0}}^{\eta} H^{(I)}[\eta,\chi^{(0)}(\eta)]\dd \eta\right)\right\}.
\end{equation} 
In this expression the operator 
 $\mT$ is the so-called "T-product", meaning that the operators that appear in its arguments should be ordered in time. Specifically the terms that appear in the argument of the exponential should each be ordered in time, e.g.,
\begin{eqnarray}
\mU^{(I)}(\eta,\eta_0)=\Id&-&\ii\int_{\eta_0}^{\eta} \dd \eta_{1}H^{(I)}(\eta_{1})\nonumber\\
&-&\int_{\eta_0}^{\eta} \dd \eta_{2}\int_{\eta_{0}}^{\eta_{2}} \dd \eta_{1}\,H^{(I)}(\eta_{2})\,H^{(I)}(\eta_{1})+\dots
\end{eqnarray}
Then using this operator one obtains for any operator  $\mQ(\eta)$, that is any function of the field $\chi(\vx,\eta)$,
\begin{equation}
\mQ(\eta)=(\mU^{(I)})^{-1}(\eta,\eta_0)\,\mQ^{(0)}(\eta)\,\mU^{(I)}(\eta,\eta_0).
\end{equation}
which again can be written,
\begin{equation}
\mQ(\eta)=\left[\overline{\mT}
\exp\left(\ii \int_{\eta_0}^{\eta} H^{(I)}\dd \eta'\right)\right]
\,\mQ^{(0)}(\eta)\,\left[{\mT}
\exp\left(-\ii \int_{\eta_0}^{\eta} H^{(I)}\dd \eta'\right)\right].
\end{equation}
This relation is the keystone of perturbation calculations for observable in an expanding universe. It implies for instance that
\begin{eqnarray}
\lvac \mQ \rvac&=&\lvac \mQ^{(0)} \rvac\nonumber\\
&&-\ii\int_{\eta_0}^{\eta}\dd \eta_{1}\,\lvac \left[\mQ^{(0)},\,H^{(I)}(\eta_{1})\right]\rvac\nonumber\\
&&-\int_{\eta_0}^{\eta}\dd \eta_{1}\int_{\eta_0}^{\eta_{1}}\dd \eta_{2}\,
\lvac \left[\left[\mQ^{(0)},\,H^{(I)}(\eta_{1})\right],H^{(I)}(\eta_{2})\right]\rvac
\nonumber\\
&&+\dots\label{QobsExp}
\end{eqnarray}
Each line of the preceding expression corresponds to a given order in perturbation theory. It is easy to see that higher order terms 
will involve expression with more and more entangled commutators.\footnote{There are alternative formulations of these expressions
in terms of time dependent propagators, see for instance \cite{2002CQGra..19.4607O,1994CQGra..11.2969T}.} 

Let us illustrate this property for observables involving simple field products and for an interaction hamiltonian that takes the fom,
\begin{equation}
H^{(I)}(\eta)\dd\eta=a^4(\eta)\dd\eta\,\int\dd^3\vx\ V\left[{\hu^{(0)}(\vx,\eta)}/{a(\eta)}\right]
\end{equation}
where $V$ is of the form of Eq. (\ref{Vshape}). In the following, we will not in particular consider terms that involve higher order derivatives of the field (although the existence of such terms cannot be excluded) and in practice we will consider terms that are cubic or quartic in the field.

As can be easily seen from the terms involved in (\ref{QobsExp}), the whole result can be written in terms of the unequal time correlation function of the  {\sl free} fields defined as,
\begin{equation}
 \mg 0\vert \hu_{\vk}(\eta) \hu_{\vk'}(\eta')\vert 0\md=(2\pi)^3
\Dirac(\vk+\vk')\ G_{k}(\eta,\eta').\label{Gdef}
\end{equation}
In can be expressed in terms of the function $u^{(0)}_{\vk}(\eta)$ that are solution of (\ref{u0evol}). And we have
\begin{equation}
G_{k}(\eta,\eta')=u^{(0)}_{\vk}(\eta)u^{(0)*}_{\vk}(\eta').
\end{equation}

In the following we explicitly give the expression of such correlation functions in case of a quartic potential. From the relation (\ref{QobsExp}) we have,
\begin{eqnarray}
\mg \chi_{\vk_1}\dots\chi_{\vk_4}\md_c&=&
\frac{1}{a^4(\eta)}\mg \hu_{\vk_1}\dots \hu_{\vk_4} \md_{c}\nonumber\\
&=&-\frac{\ii}{a^4(\eta)}\int^{\eta}\dd\eta' \mg 0\vert
  \left[u_{\vk_{1}}^{(0)}(\eta)\dots u_{\vk_{4}}^{(0)}(\eta),H_I(\eta')\right]\vert 0\md\nonumber\\
  &=&-\frac{\ii\lambda}{4!\,a^4(\eta)}\int^\eta\dd\eta'\frac{\dd^3\vx'}{(2\pi)^{12}}{\dd\vk'_{1}}\dots {\dd\vk'_{4}}\ e^{\ii\vx'.\sum_{i}\vk'_{i}}\nonumber\\
 &\times& \mg 0\vert\!  \left[\hu_{\vk_{1}}^{(0)}(\eta)\dots \hu_{\vk_{4}}^{(0)}(\eta),\hu_{\vk'_{1}}^{(0)}(\eta')\dots \hu_{\vk'_{4}}^{(0)}(\eta')\right]\!\vert 0\md
\end{eqnarray}
It finally leads to,
\begin{eqnarray}
\mg \chi_{\vk_1}\dots\chi_{\vk_4}\md_c&=&-\frac{\ii\lambda}{a^4(\eta)}(2\pi)^3\Dirac\left(\sum_{i=1}^{4}\vk_{i}\right)\nonumber\\
&&\times\int^{\eta}{\dd\eta'}\left[G_{k_{1}}(\eta,\eta')\dots G_{k_{4}}(\eta,\eta')-\hbox{c.c.}\right]
\label{fourthcumgeneral}
\end{eqnarray}
where $\hbox{c.c.}$ stands for the complex conjugate of the previous term. Note that from the definition of $G_{k}(\eta,\eta')$ in Eq. (\ref{Gdef}) it is clear that $G_{k}(\eta',\eta)$ is the complex conjugate of $G_{k}(\eta,\eta')$. In practice the computation of such cumulants relies therefore on our knowledge of the explicit form of the $G$ functions. In the following we present two specific cases where the calculations can be completed analytically. These are the cases of de Sitter background and some models of power law inflation.

\subsection{Three- and Four-point function in de Sitter background}

For a de Sitter background we have $a''(\eta)/a(\eta)=-2/\eta^2$ and the motion equation of Eq. (\ref{u0evol}) can be solved exactly. We then have
\begin{equation}
u_{k}^{(0)}(\eta)=\frac{1}{\sqrt{2k}}\left(1-\frac{\ii}{k\eta}\right)e^{-\ii k \eta}.
\end{equation}
%and therefore,
%\begin{equation}\label{dSpowerspectrum}
%  G_{k}^{\rm dS}(\eta,\eta')=\frac{1}{2k}\left(1-\frac{\ii}{k\eta}\right)
%  \left(1+\frac{\ii}{k\eta'}\right)\exp[\ii k(\eta'-\eta)]
%\end{equation}
It is interesting to note that this solution can be rewritten,
\begin{equation}
u_{k}^{(0)}(\eta)=\frac{1}{\sqrt{2k}\,\eta}\mO_{k}\,e^{-\ii k \eta}
\end{equation}
where the operator $\mO_{k}$ is defined as 
\begin{equation}
\mO_{k}=\frac{1}{k}-\frac{\partial}{\partial k}.
\end{equation}
As a result, in this case, the expression (\ref{fourthcumgeneral}) can be rewritten,
\begin{eqnarray}
\mg \chi_{\vk_1}\dots\chi_{\vk_4}\md_c&=&-\frac{\ii\lambda}{16\pi_{4}\eta^4a^4(\eta)}(2\pi)^3\Dirac\left(\sum_{i=1}^{4}\vk_{i}\right)
\nonumber\\
&&\hspace{-3cm}\times\left\{
\bigg(\mO_{k_{1}}\dots\mO_{k_{4}}\ e^{-\ii \pi_{1}\eta}\bigg)
\left(\mO_{k_{1}}\dots\mO_{k_{4}}\int^{\eta}\frac{\dd\eta'}{\eta'^4}\ e^{\ii \pi_{1}\eta'}\right)-c.c.
\right\}\label{fourthcumdS}
\end{eqnarray}
where we introduce the following combination of the four wave modes,
\begin{eqnarray}
\pi_1&=\sum_i k_i ,\ \ 
&\pi_2=\sum_{i < j} k_i\,k_j ,\nonumber\\ 
\pi_3&=\sum_{i < j < k} k_i\,k_j\,k_k ,\ \ 
&\pi_4=\sum_{i < j < k < l} k_i\,k_j\,k_k\,k_l.
\end{eqnarray}
The integral appearing in Ref. (\ref{fourthcumdS}) reads 
\begin{equation}
\int^{\eta}\frac{\dd\eta'}{\eta'^4}\ e^{\ii \pi_{1}\eta'}=
-\frac{1}{6} \ii\ \textEi\left(\ii \eta  \pi _1\right) \pi _1^3+\frac{e^{\ii \eta  \pi _1} \pi _1^2}{6 \eta }-\frac{\ii e^{\ii \eta  \pi _1} \pi _1}{6 \eta ^2}-\frac{e^{\ii \eta  \pi _1}}{3 \eta ^3}
\end{equation}
from which the explicit expression of $\mg \chi_{\vk_1}\dots\chi_{\vk_4}\md_c$ can easily be derived (it is given in Ref. \cite{2004PhRvD..69f3520B} in terms of the combinations $\pi_{i}$).

%\begin{figure}[htbp]
%\begin{center}
%\epsfig{file=deSitterzetatot.eps,scale=.7}
%\caption{Dependence of $\zeta$ on $k_4$ for fixed values of $k_1$, $k_2$ and $k_3$. 
%The continuous line corresponds to the configuration  $k_1$=0, $k_2$=$\frac{1}{2}$,
%$k_3$=$\frac{1}{2}$, $k_4$ ; the dashed line to
%$k_1$=$\frac{1}{3}$, $k_2$=$\frac{1}{3}$, $k_3$=$\frac{1}{3}$,
%$k_4$ and the dotted line to  $k_1$=$\frac{1}{6}$,
%$k_2$=$\frac{1}{3}$, $k_3$=$\frac{1}{6}$, $k_4$.}
%\label{zetatot}
%\end{center}
%\end{figure}

In the the super-Hubble limit (i.e.  $k_i\eta\ll 1$ for all $i=1\ldots4$) the expression of $Q_{4}$ as defined in Eq. (\ref{Qpdef}) can be  explicitly computed giving,\begin{equation}
Q_{4}(\left\{k_{i}\right\}) = \frac{\lambda
}{3 H^2}
\left[\gamma+\zeta(\{k_i\})+\log\left(-\eta\sum k_i\right)\right].\label{fourthcumlimit}
\end{equation}
In this expression, terms of order $k_i\eta$ have been neglected. The $\zeta$ function is an homogeneous function of the wave vectors
which reads, 
\begin{equation}\label{zeta}
  \zeta(\{k_i\})=\frac{-{\pi_1}^4 +
2\,{\pi_1}^2\,\pi_2 + \pi_1\,\pi_3 - 3\,\pi_4}{\pi_1\,\left(
{\pi_1}^3 - 3\,\pi_1\,\pi_2 + 3\,\pi_3 \right)}.
\end{equation}
It depends on the wave vector ratios but in a rather weak way. Its minimum value, $-51/16$, is reached for a symmetric configuration of the 4 wave modes. Its maximum, $-2$, is reached when 2 of the wave mode values vanish.

Obviously similar results can be obtained for the 3-point function in case of a cubic potential. In this case the quantity to compute is 
\begin{eqnarray}
\mg \chi_{\vk_1}\dots\chi_{\vk_3}\md&=&-\frac{\ii\lambda}{8\pi_{3}\eta^3 a^3(\eta)}(2\pi)^3\Dirac\left(\sum_{i=1,3}\vk_{i}\right)
\nonumber\\
&&\hspace{-3cm}\times\left\{
\bigg(\mO_{k_{1}}\dots\mO_{k_{3}}\ e^{-\ii \pi_{1}\eta}\bigg)
\left(\mO_{k_{1}}\dots\mO_{k_{3}}\int^{\eta}\frac{\dd\eta'}{\eta'^4}\ e^{\ii \pi_{1}\eta'}\right)-c.c.
\right\}.\label{thirdcumdS}
\end{eqnarray}
Interestingly, to a factor $2\eta$, one recovers the same expression as for the four-point function when $\pi_{4}$ is put to zero so that $Q_{3}$, once again defined from Eq. (\ref{Qpdef}),
takes the form,
\begin{equation}
Q_{3} = \frac{\lambda}{3 H^2}\left[\gamma+\zeta_3(\{k_i\})+\log\left(-\eta\sum k_i\right) \right]\label{thirdcumlimit}
\end{equation}
in the super-horizon limit where $\zeta_3$ is,
\begin{equation}\label{zeta3exp}
  \zeta_3(\{k_i\})=\zeta(k_1,k_2,k_3,k_4\to 0).
\end{equation}
\begin{figure}[htbp]
\begin{center}
\epsfig{file=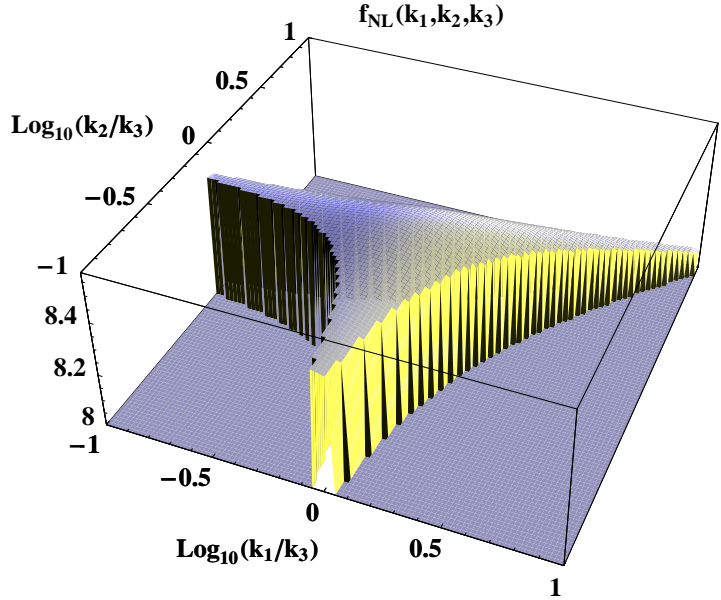,scale=.7}
\caption{Dependence of $f^{\chi}_{\rm NL}$ on $k_i$ as a function of $k_{1}$ and $k_{2}$ for a fixed value of $k_{3}$ and for $\lambda/H^2=1$. This function is computed assuming the numbers of efolds since $k_{3}$ crossed the horizon is $N_{e}=50$. }
\label{Q3dS}
\end{center}
\end{figure}
%This corresponds to the solid line on Fig. \ref{zetatot}. 
The values of $\zeta_3(\{k_i\})$ range from $-8/3$ to $-2$.
This expression gives the $k_{i}$ dependence of the usual  $f^{\chi}_{\rm NL}$ parameter for the $\chi$ field -- which is nothing but $Q_{3}/2$) -- as,
\begin{eqnarray}
f^{\chi}_{\rm NL}(\{k_i\})&=&-\frac{\lambda}{6 H^2}\left[N_{e}(k_{3})-\log\left(\frac{k_{1}}{k_{3}}+\frac{k_{2}}{k_{3}}\right) -\gamma\right.
\nonumber\\
&&\left.-\frac{k_1 k_2 k_3-\left(k_1+k_2+k_3\right) \left(k_1^2+k_2^2+k_3^2\right)}{k_1^3+k_2^3+k_3^3}\right],\label{fnllimit}
\end{eqnarray}
were $N_{e}(k_{3})$ is the number of efolds between the horizon crossing of $k_{3}$ and current time. A similar result can be derived for $\tau_{\rm NL}^{\chi}=Q_{4}/6$. They both scale like $\lambda N_{e}/H^2$ and therefore give an order of unity parameters for
$\lambda N_{e}\approx H^2$ which would then be an effect comparable to the gravity induced couplings. Those results are therefore
valid, and of interest, for $H^2\lesssim \lambda N_{e}\lesssim H$ for a cubic coupling and $H^2\lesssim \lambda N_{e}\lesssim 1$ for a quartic coupling.

%%%%%%

The resulting $k-$dependence of $f^{\chi}_{\rm NL}$ is shown on Fig. \ref{Q3dS} for a $N_{e}(k_{3})=50$ and $\lambda/H^2=1$. What this figure shows is the remaining dependence of the non-linear coefficient on configurations. Clearly the main feature is that $f_{\rm NL}$ is almost independent of the configuration and of scale. To be more precise it is straightforward to see that, for a given configuration (e.g. given ratios of $k_{i}$),
\begin{equation}
\frac{\dd \log f^{\chi}_{\rm NL}}{\dd \log k_{t}}=\frac{1}{N_{e}(k_{t})}
\end{equation}
where $k_{t}=k_{1}+k_{2}+k_{3}$ and $N_{e}$ is the number of efolds since $k_{t}$ crossing. We are thus left with a 2 percent effects. On Fig. \ref{Q3dS} one can see however that the dependence of $f^{\chi}_{\rm NL}$ on the geometry can be as large as $7.5$ percent for wave mode ratio of $10$. These dependences are entirely encoded in the form of Eq. (\ref{fnllimit}). In the following we will see how they depend on the background dynamics.

%What we want to explore in the following is the robustness of the result when the background is changed.
%In both cases what we will show here is that the leading $\log$ in (\ref{fourthcumlimit}) and in (\ref{thirdcumlimit}) corresponds to the classical behavior, whereas the sub-leading terms are genuinely of quantum origin. The aim of the paper is precisely to explicit the connection between the leading behavior of the vertices and the background evolution.

\subsection{A case of power-law inflation}
There is another simple (yet unrealistic) model of inflation which can be solved explicitly and that corresponds to a case of power-law inflation. In the latter case the expansion factor is assumed to behave like $t^{\nu}$ with $\nu > 1$
where $t$ is the physical time or equivalently $a(\eta)\sim \eta^{-\nu/(\nu-1)}$. In such cases the evolution equation for $u_{k}^{(0)}(\eta)$
can be solved in
\begin{equation}
u_{k}^{(0)}(\eta)=\frac{\sqrt{-\pi\eta}}{2}H^{(2)}_{\mu}(-k\eta)
\end{equation}
where $H^{(2)}_{\mu}$ is the function of Hankel of the second kind with
\begin{equation}
\mu=\frac{3}{2}+\frac{1}{\nu-1}.
\end{equation}
When $\mu$ is a half integer it is possible to explicitly compute the expression of the four-point function of $\chi$. Besides the de Sitter case (that corresponds to $\mu=3/2$), the case\footnote{There are actually an infinite number of cases that could be solved following this procedure but they would correspond to smaller values of $\nu$.} $\nu=2$ leads to $\mu=5/2$ and for which we have
\begin{equation}
u_{k}^{(0)}(\eta)=\frac{1}{\sqrt{2k}\,\eta^2}\,\mOd_{k}\,e^{-\ii k \eta}
\end{equation}
 with
 \begin{equation}
\mOd_{k}=\frac{3}{k^2}-\frac{3}{k}\frac{\partial}{\partial k}+\frac{\partial^2}{\partial k^2}.
\end{equation}
In this case the late time power spectrum reads
\begin{equation}
P_{\chi}(k)=\frac{9}{2}\frac{1}{k^5\ \eta^4 a^2(\eta)}=\frac{9}{8}\frac{k_{0}^2H_{0}^2}{k^5}
\end{equation}
where $H_{0}$ is the value of $H$ at time of horizon crossing for the mode $k_{0}$.
The formal expression of high-order terms can also be computed. For a quartic potential it leads to the following form,
\begin{eqnarray}
\mg \chi_{\vk_1}\dots\chi_{\vk_4}\md_c&=&-\frac{\ii\lambda}{16\pi_{4}\,\eta^4\,a^4(\eta)}(2\pi)^3\Dirac\left(\sum_{i=1,4}\vk_{i}\right)
\nonumber\\
&&\hspace{-3cm}\times\left\{
\bigg(\mOd_{k_{1}}\dots\mOd_{k_{4}}\ e^{-\ii \pi_{1}\eta}\bigg)
\left(\mOd_{k_{1}}\dots\mOd_{k_{4}}\int^{\eta}\frac{\dd\eta'}{\eta'^8}\ e^{\ii \pi_{1}\eta'}\right)-c.c.
\right\}.\label{fourthcumPL}
\end{eqnarray}
The super-Hubble limit of this expression can again be computed at leading orders in super-horizon. It leads to
\begin{eqnarray}
Q_{4}\left(\left\{k_{i}\right\}\right)&=&\frac{\lambda\ a^2(\eta)\eta^2}{14}\nonumber\\
&&\hspace{-.5cm}\times\left[
1+\frac{\eta^2\left(3\sum k_{i}^7-7\sum_{j\ne j}k_{i}^5 k_{j}^2\right)}{15\sum k_{i}^5}\log(-\eta \sum_{i}k_{i})
\right].
\end{eqnarray}
The growth of the amplitude of the reduced correlation function is then extremely rapid since it grows like $a(\eta)$, that is like $\exp(N_{e})$ where $N_{e}$ is the number of efolds. The corrective term is only in $N_{e}$. This is clearly at variance with the 
result in case of the de Sitter background case. That implies in particular that the $k$ dependence of the $f_{\rm NL}^{\chi}$ and $\tau_{NL}^{\chi}$ parameters is completely suppressed, by many orders of magnitude.

In the following we aim at exploring in a systematic way the dependence of the late time behavior of the function $Q_{p}$ on the background dynamics. It will not be possible however to do so from exact quantum calculations with the ``in-in'' formalism and we rather take advantage of the classical re-formulation of the mode coupling evolution for super-horizon scales.

\section{Quantum to classical descriptions}

The classical system we introduce now is not exactly formulated as in Ref. \cite{2008PhRvD..78f3534W} . It will nonetheless exhibit the same super-Hubble properties.

\subsection{A classical stochastic counterpart system}

So let us now introduce a classical system that consists in a stochastic real field, $\cu(\vx,\eta)$, the time evolution of which is given by
Eq.  (\ref{uevol1}). As a classical field, the amplitude 
of $\cu(\vx,\eta)$ is determined by its initial conditions that we take corresponding to those of the free quantum field $\hu^{(0)}(\vx,\eta)$.

Remark that if we define
\begin{equation}
\hs_{\vk}=\left(a_{\vk}+\ad_{-\vk}\right)\ \ \hbox{and}\ \ \hd_{\vk}=\left(a_{\vk}-\ad_{-\vk}\right)
\end{equation}
then the relation (\ref{u0exp1}) can be rewritten
\begin{equation}
\hu^{(0)}(\vx,\eta)=\int\frac{\dd^3\vk}{(2\pi)^3} e^{\ii\vk.\vx}\left[{\hbox{Re}(u^{(0)}_{\vk}(\eta))}\,\hs_{\vk}+{\hbox{Im}(u^{(0)}_{\vk}(\eta))}\,\hd_{\vk}\right].\label{u0exp2}
\end{equation}
We can then note that
\begin{equation}
\mg 0\vert \hs_{\vk}\hs_{\vk'} \vert 0\md=\mg 0\vert \hd_{\vk} \hd_{\vk'} \vert 0\md=\hbar\,(2\pi)^3\Dirac(\vk+\vk').
\end{equation}
For the classical model, the free field $\cu^{(0)}$ obeys Eq. (\ref{u0exp2}) when $\hs$ and $\hd$ are ``degraded'' into time independent classical random variables that form each  a Gaussian field.  More precisely we introduce the random variables $\cs$ and $\cd$ as time \textsl{independent} Gaussian random variables whose ensemble averages are given by,
\begin{equation}
\mg \cs_{\vk}\cs_{\vk'} \md=\mg  \cd_{\vk}\cd_{\vk'} \md=\hbar\,(2\pi)^3\Dirac(\vk+\vk'),
\end{equation}
so that 
\begin{eqnarray}
\cu^{(0)}(\vx,\eta)&=&\int\!\frac{\dd^3\vk}{(2\pi)^3} e^{\ii\vk.\vx}\left[{\hbox{Re}(u^{(0)}_{\vk}(\eta))}\,\cs_{\vk}+{\hbox{Im}(u^{(0)}_{\vk}(\eta))}\,\cd_{\vk}\right]\label{u0exp2c}
\end{eqnarray}
is a classical free field the evolution of which is fully compatible with  (\ref{u0evol}) (this is clearly the case since both $\hbox{Re}(u^{(0)}_{\vk}(\eta))$
and $\hbox{Im}(u^{(0)}_{\vk}(\eta))$ are solutions of the linearized field equations).
As a result the classical modes, $\cu^{(0)}_{\vk}(\eta)={\hbox{Re}(u^{(0)}_{\vk}(\eta))}\,\cs_{\vk}+{\hbox{Im}(u^{(0)}_{\vk}(\eta))}\,\cd_{\vk}$,
obey the following properties,
\begin{equation}
\mg\cu^{(0)}_{\vk}(\eta)\cu^{(0)}_{\vk'}(\eta')\md=(2\pi)^3\Dirac(\vk+\vk')\,P_{k}(\eta,\eta')
\end{equation}
with
\begin{eqnarray}
P_{k}(\eta,\eta')&=&\hbox{Re}(u^{(0)}_{\vk}(\eta))\hbox{Re}(u^{(0)}_{\vk}(\eta'))+\hbox{Im}(u^{(0)}_{\vk}(\eta))\hbox{Im}(u^{(0)}_{\vk}(\eta'))\nonumber\\
&=&\hbox{Re}(u^{(0)}_{\vk}(\eta)u^{(0)*}_{\vk}(\eta'))
\end{eqnarray}
which is nothing but the real part of $G_{k}(\eta,\eta')$. 

The connection between the two systems is actually more profound and extends to non-linear evolution properties. It can be easily checked that the imaginary part of $G_{k}(\eta,\eta')$ is -- to a factor 2 -- the Green function of the classical system. The classical Green function $G^{C}(\eta,\eta')$ can be written
\begin{equation}
G^C_{\vk}(\eta,\eta')=\Theta(\eta-\eta')g_{\vk}(\eta,\eta')
\end{equation}
where $\Theta$ is the Heavyside function and $g_{\vk}(\eta,\eta')$ can be constructed from any two solutions of Eq. (\ref{u0exp2}).
It can easily be shown that, given the normalization constraints on $u_{\vk}^{(0)}(\eta)$, we have
\begin{equation}
G_{k}(\eta,\eta')=P_{k}(\eta,\eta')-\frac{\ii}{2} g_{k}(\eta,\eta').
\end{equation}
It gives some direct insights into the transition of the quantum evolution into the classical regime. Indeed, 
at super-horizon scales, the phase of $u^{(0)}_{k}(\eta)$ tends to be fixed which implies that the imaginary part of the $G_{k}(\eta,\eta')$ functions decays with respect to its real part. As we see here this behavior is closely related to the classical evolution of the field for super-Hubble scales since $G_{k}(\eta,\eta')$ then tends to be simply the classical unequal time field correlation function.

This observation has also some consequences on the expression of the r.h.s of Eq. (\ref{fourthcumgeneral}). Indeed it can be noted that as one takes the imaginary part of a product of $G_{k}$ functions, the result necessarily involves at least one $g_{k}$ factor.  In other words, the expression  (\ref{fourthcumgeneral}) can be rewritten,
\begin{eqnarray}\label{Q4ptsExp}
\mg \chi_{\vk_1}\dots\chi_{\vk_4}\md_c&=&-\frac{\lambda}{a^4(\eta)}\Dirac\left(\sum_{i=1}^{4}\vk_{i}\right)
\nonumber\\
&&\times\int^{\eta}{\dd\eta'}\left[ P_{k_{1}}(\eta,\eta')\dots P_{k_{3}}(\eta,\eta')g_{k_{4}}(\eta',\eta)\right.\nonumber\\
&&\left.-\frac{1}{8} P_{k_{1}}(\eta,\eta')g_{k_{2}}(\eta,\eta')\dots g_{k_{4}}(\eta,\eta')+\hbox{sym.}\right].
\end{eqnarray}
Naturally the first term appearing  in the bracket is to be the leading contribution for super-horizon scales and it will turn out with a simple classical interpretation.

\subsection{Classical evolution of the high-order correlation functions}

We focus here on the mode coupling effects induced by simple forms for the potential, that is
$\mg u_{\vk_{1}}\dots u_{\vk_{3}}\md$ in case of a cubic coupling and $\mg u_{\vk_{1}}\dots u_{\vk_{4}}\md_{c}$ in case of a quartic coupling. In order to be able to apply a perturbation theory approach we have to assume that $u_{\vk}$ can be expanded in terms of the coupling constant,
\begin{equation}
u_{\vk}(\eta)=u_{\vk}^{(0)}(\eta)+u_{\vk}^{(1)}(\eta)+\dots\label{uexpansion}
\end{equation}
where $u^{(n)}$ is of order $n$ in the coupling constant. The first order expression of $u$ can easily by computed from the classical 
motion equation, e.g. we have,
\begin{equation}
{u^{(1)}_{\vk}}''(\eta)-\frac{a''}{a}u^{(1)}_{k}(\eta)+k^2\,u^{(1)}_{k}(\eta)=-a^3(\eta)
\int {\dd^3\vx}\frac{\dd V}{\dd\chi}\,e^{-\ii\vk.\vx}
\end{equation}
where $\frac{\dd V}{\dd\chi}$ is the derivative of the potential with respect to $\chi$ and taken at value ${u^{(0)}(\vx,\eta)}/{a(\eta)}$. The solution of this motion equation can be written with the help of the classical Green function, $g_{k}(\eta,\eta')$ as,
\begin{equation}
{u^{(1)}_{\vk}}(\eta)=-\int^{\eta}a^3(\eta')g_{k}(\eta,\eta')
\int {\dd^3\vx}\exp\left({-\ii\vk.\vx}\right)\,\frac{\dd V}{\dd\chi}.
\end{equation}
In the following we treat the case of a quartic\footnote{the case of a cubic coupling can be treated very similarly.} coupling $V(\chi)=\frac{\lambda}{4!}\chi^4$. 
In this case, the double Fourier transforms lead to,
\begin{eqnarray}
{u^{(1)}_{\vk}}(\eta)&=&-\frac{\lambda}{3!(2\pi)^6}\int^{\eta}g_{k}(\eta,\eta')
\int\dd^3\vk_{1}\dots \dd^3\vk_{3}\nonumber\\
&&\times u^{(0)}_{\vk_{1}}(\eta')\dots u^{(0)}_{\vk_{3}}(\eta')\,
\Dirac\left(\vk-\sum_{i=1}^3\vk_{i}\right).
\end{eqnarray}
As a result, the tree order contribution of the fourth order cumulant reads,
\begin{eqnarray}
\mg\chi_{\vk_{1}}\dots\chi_{\vk_{4}}\md_{c}
&=&-\frac{\lambda(2\pi)^3}{a^4(\eta)}\Dirac\left(\sum_{i=1}^{4}\vk_{i}\right)\nonumber\\
&&\hspace{-2cm}\times\int_{\eta_{0}}^{\eta}{\dd\eta'}\ P_{k_{1}}(\eta,\eta')P_{k_{2}}(\eta,\eta')P_{k_{3}}(\eta,\eta')\ g_{k_{4}}(\eta,\eta')+\hbox{sym.}
\label{chi4exp2}
\end{eqnarray}
This expression reproduces Eq. (\ref{Q4ptsExp}) when the last contribution has been dropped. At variance with the classical system proposed in Ref. \cite{2008PhRvD..78f3534W}, the evolutions of the two systems are here not identical, even at tree order. However, provided the integral appearing in this expression is
dominated by super-Hubble contributions -- and this is indeed the case for the two examples we know how to compute -- so that
$g_{k}(\eta,\eta')$ is much smaller than $P_{k}(\eta,\eta')$, the two contributions should be nearly equal. This is this observation that allows us to sort out the leading contributing terms to such correlation functions.

%

%\begin{figure}[htbp]
%\begin{center}
%\epsfig{file=NuEff3Equi.eps,width=8cm}
%\epsfig{file=NuEff3Squeezed.eps,width=8cm}
%\caption{The effective vertex for the three-point function as a function of the number of efolds since horizon crossing for a de Sitter background. The left panel corresponds to an equilateral configuration ($k_{1}=k_{2}=k_{3}$) and the right panel to a squeezed
%configuration ($k_{2}=k_{3}=10.\,k_{1}$).}
%\label{NuEff3}
%\end{center}
%\end{figure}

%
%\begin{figure}[htbp]
%\begin{center}
%\epsfig{file=NuEff4Equi.eps,width=8cm}
%\epsfig{file=NuEff4Squeezed.eps,width=8cm}
%\caption{The effective vertex for the four-point function as a function of the number of efolds since horizon crossing. The left panel corresponds to an equilateral configuration ($k_{1}=k_{2}=k_{3}=k_{4}$) and the right panel to a squeezed
%configuration ($k_{2}=k_{3}=10.\,k_{1}=10.\,k_{2}$).}
%\label{NuEff4}
%\end{center}
%\end{figure}

%
%At this stage it is possible to compare the results of the quantum calculations to the classical calculations. This comparison is shown on Figs.
%\ref{NuEff3} and \ref{NuEff4}. As expected they exhibit similar behavior; differ in general but leads to the same asymptotic behavior. The abscissa is in units of $N_{e}$, computed from the time of horizon crossing, and it is clear that the effective vertex grows linearly in $N_{e}$ in all cases. What the figure show however is that there are finite differences in the vertex for different configurations. Those differences are specifically of quantum origin. They cannot be reproduced in the classical calculation.

%

\section{Dependence on the background dynamics}

\subsection{General expressions}

The practical computation of the leading behavior of Eq. (\ref{chi4exp2}) can then be further simplified by observing that
one can use  the classical Green function of the system in its limit\footnote{It is to be noted that the sub-dominant terms cannot be accounted for in a perturbative expansion of $g$ with respect to $k\eta$. For instance next-to-leading order in de Sitter case can only be derived from a full use of Eq. (\ref{Q4ptsExp}).} $k\to 0$. 
Furthermore, for superhorizon scales, the dependence of $P_{k}(\eta,\eta')$ on $\eta'$ is that of $a(\eta')$. 
The expression (\ref{chi4exp2}) is then simply proportional to,
\begin{equation}
\nue(\eta)=\frac{\lambda}{a(\eta)}\int\dd\eta'\,a(\eta')^3\,g_{0}(\eta,\eta')\label{nueexp}
\end{equation}
which is nothing but the super-Hubble limit of $Q_{4}$,
\begin{equation}
Q_{4}(\eta,\{k_{i}\})\to \nue(\eta).
\end{equation}
It is interesting to note that a calculation with a cubic potential would actually have led to the very same expression for $Q_{3}(\eta,\{k_{i}\})$. In the following we will then note $\nue(\eta)$ the common values of $Q_{p}(\eta)$ in the super-Hubble limit.

We now turn to the explicit computation of $g_{0}(\eta,\eta')$.
As we are considering massless fields, it can easily be derived from the behavior of the expansion factor. One solution of the field evolution is indeed given by, $v_{1}(\eta)=a(\eta)$, the other solution is given by 
$
v_{2}(\eta)=a(\eta)\int^{\eta} {\dd \eta'}/{a^2(\eta')},
$
so that the classical Green function then reads $g_{0}(\eta,\eta')\Theta(\eta-\eta')$ in the $k\to 0$ limit with,
\begin{equation}
g_{0}(\eta,\eta')=a(\eta)a(\eta')\,\int_{\eta'}^{\eta}\frac{\dd\eta''}{a^2(\eta'')}.\label{g0exp}
\end{equation}
The expression (\ref{nueexp}) can then be formally written,
\begin{equation}
\nue(\eta)={\lambda}\int_{\eta_{0}}^{\eta}\dd\eta'\,a(\eta')^{4}\,\int_{\eta'}^{\eta}\frac{\dd \eta''}{a^2(\eta'')},\label{nueexpression}
\end{equation}
in terms of the expansion factor. This is one important result of this paper.

In the following we are going to explore the dependence of this quantity with the background dynamics. Let us start with the de Sitter case for which we have, 
\begin{equation}
g_{0}^{\rm dS}(\eta,\eta')=\frac{1}{3\eta\eta'}\left(\eta^3-\eta'^3\right)
\end{equation}
which is indeed what the late time expansion of the imaginary part of $G^{\rm dS}$ gives. We can now plug in this behavior into
(\ref{nueexp}). In this case we then have,
\begin{equation}
\nue^{\rm dS}(\eta)=\frac{\lambda}{H^2}\int_{\eta_{0}}^{\eta}\frac{\dd\eta'}{3\eta'^4}\left(\eta^3-\eta'^3\right)=\frac{\lambda}{H^2}
\left(\frac{\eta^3}{\eta_{0}^3}-1-\frac{\log(\eta/\eta_{0})}{3}\right)
\end{equation}
During the super-horizon evolution $\eta$ gets close to $-0$ and $\eta/\eta_{0}$ gets very small. The number of efolds after time $\eta_{0}$ is nothing but $N_{e}=-\log(\eta/\eta_{0})$. It corresponds, to a finite correction, to the number of efolds after horizon crossing, e.g. the validity domain of the form (\ref{g0exp}) of $g_{k}(\eta,\eta')$. In case of a de Sitter background the vertex value is then
$\nue^{\rm dS}(\eta)={\lambda N_{e}}/{3H^2}$.

\subsection{Power-law inflation}

We can now return to the power law case but without any restriction on the index. So we assume that the expansion factor behaves like $t^{\nu}$ with $\nu > 1$ where $t$ is the physical time. In this case, we have $H=\nu/t$ and
\begin{equation}
g_{0}^{\rm pl}(t,t')=t^{\nu}t'^{\nu}\,\int_{t'}^{t}\frac{\dd t'}{t'^{3\nu}}=-
\frac{t^{\nu}\,t'^{\nu}}{3\nu-1}\left(t^{1-3\nu}-t'^{1-3\nu}\right)
\end{equation}
which implies that
\begin{eqnarray}
\nue^{\rm pl}(\eta)&=&-{\lambda}\int_{t_{0}}^{t}\dd t'\,\frac{t'^{3\nu}}{3\nu-1}\left(t^{1-3\nu}-t'^{1-3\nu}\right)\\
&=&\frac{\lambda}{3\nu-1}\left(\frac{t^2-t_{0}^2}{2}-\frac{t^2-t_{0}^{3\nu+1}/t^{3\nu-1}}{3\nu+1}\right).
\end{eqnarray}
For $\nu=2$ one recovers the late time behavior found in the exact calculation. In general 
this expression can be reformulated in terms of $N_{e}$ and $H_{0}$, the value of $H$ at horizon crossing time $t_{0}$. We have
$t/t_{0}=\exp(N_{e}/\nu)$ so that,
\begin{eqnarray}
\nue^{\rm pl}(\eta)&=&\frac{\lambda \nu^2}{(3\nu-1)H_{0}^2}\nonumber\\
&&\times\left[
\frac{1}{2}\left(e^{2N_{e}/\nu}-1\right)-\frac{1}{3\nu+1}\left(e^{2N_{e}/\nu}-e^{(1-3\nu)N_{e}/\nu}\right)\right].\label{nueExpPL}
\end{eqnarray}
If one further assumes that $\nu/N_{e}$ is large enough, then we have
\begin{equation}
N_{e}=\log\frac{a(t)}{a(t_{0})}=\log\left(\frac{t}{t_{0}}\right)^{\nu}\approx \nu \frac{t-t_{0}}{t_{0}}
\end{equation}
so that
\begin{equation}
\nue^{\rm pl}(\eta)\approx\frac{\lambda}{3\nu}(t-t_{0})t_{0}\approx\frac{\lambda N_{e}}{3H_{0}^2},
\end{equation}
which reproduces the result of the de Sitter case. Note however that this convergence is very poor. A simple expansion 
of (\ref{nueExpPL}) gives
\begin{equation}
\nue^{\rm pl}(\eta)\approx\frac{\lambda N_{e}}{3H_{0}^2} \left(1+\frac{N_{e}}{3\nu}+\frac{2N_{e}^2}{3\nu^2}+\dots\right).
\end{equation}

\begin{figure}[htbp]
\begin{center}
\epsfig{file=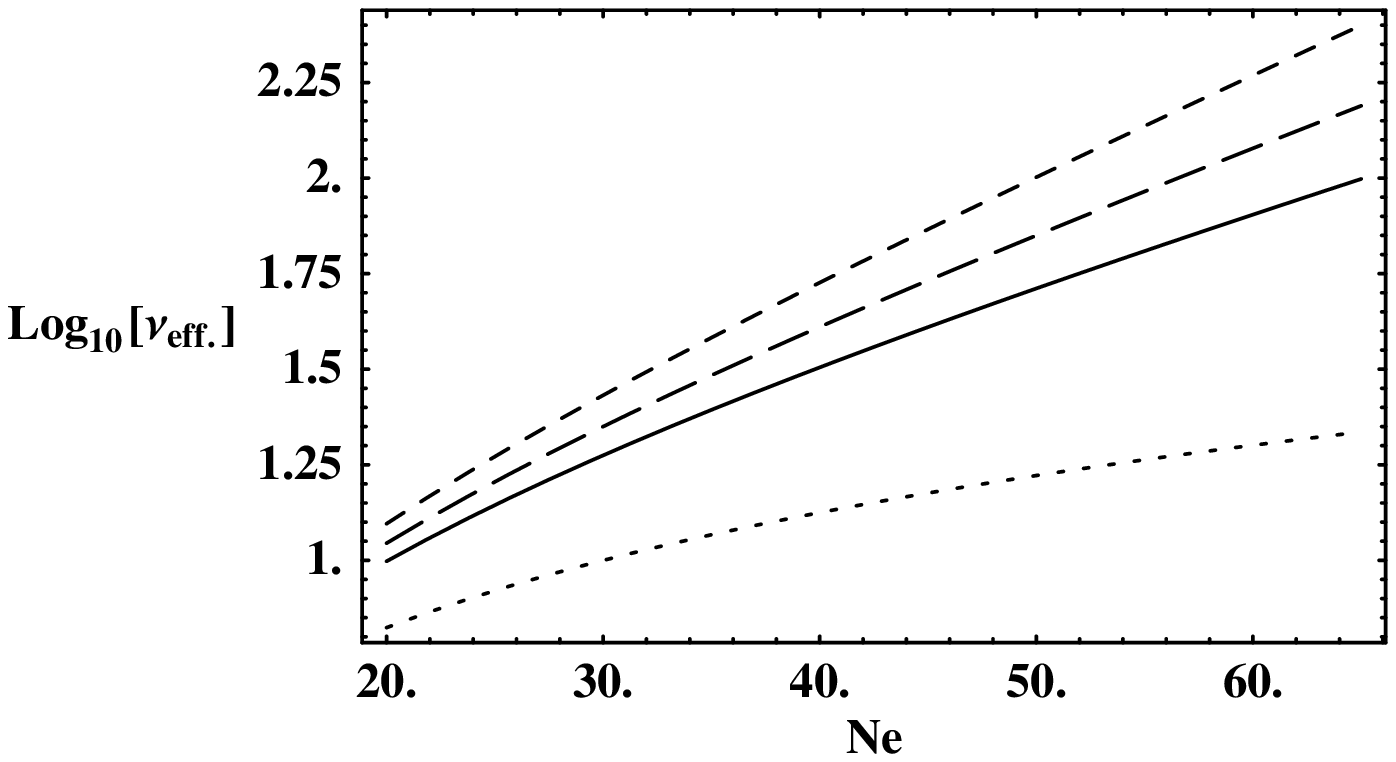,width=8cm}
\epsfig{file=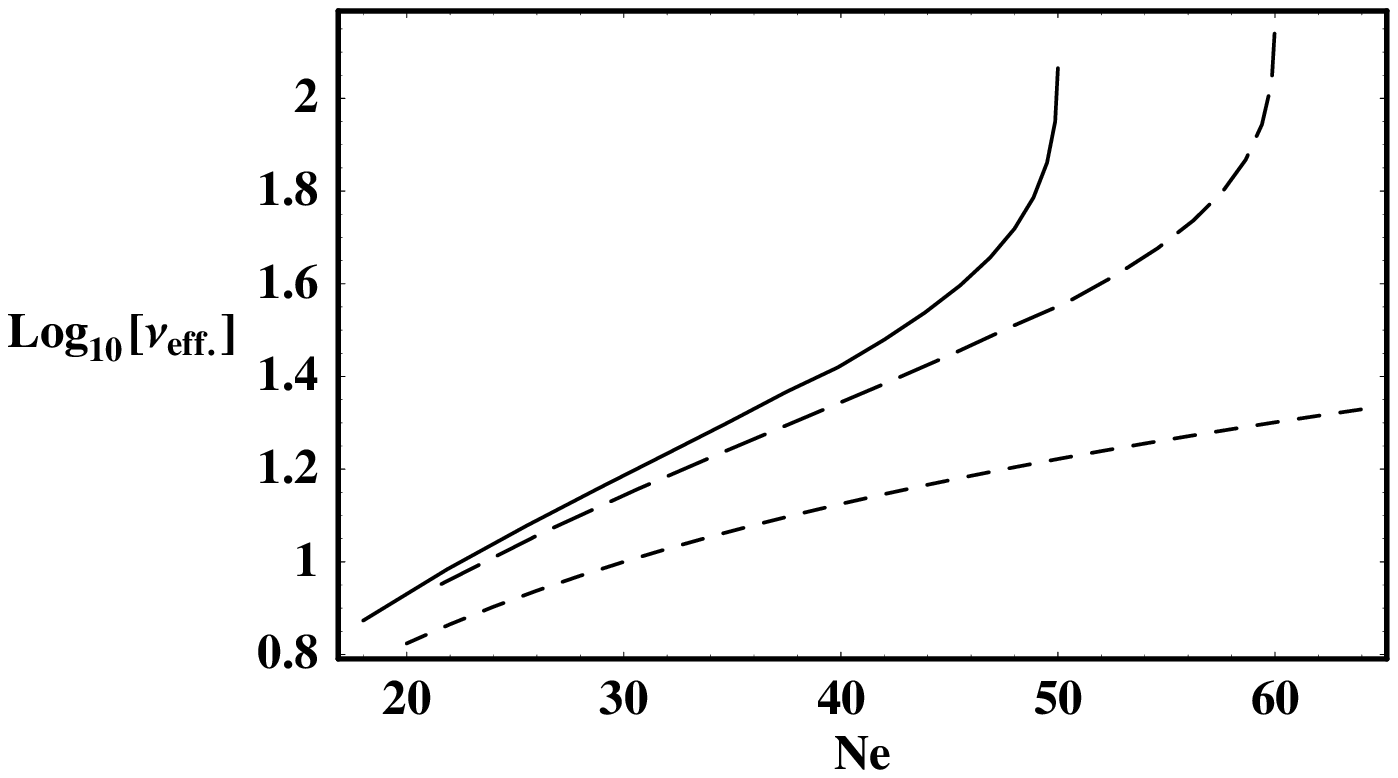,width=8cm}
\caption{Behavior of the effective vertex value as a function of the number of efolds from horizon crossing. The value of the vertex is given in Log scale in units such that $\lambda/H_{0}^2=1$. \\
Top panel: power law inflation. The solid line corresponds to $\nu=50.$; the long dashed line to $\nu=41$ and the short dashed line to $\nu=35.$ The dotted line to the de Sitter result.\\
Bottom panel: inflation from a massive field. The solid line corresponds to a case where the total number of efolds is $50$ whereas the long dashed line corresponds to $N_{e}=60$. The short dashed line corresponds to the de Sitter background case. }
\label{NuEff}
\end{center}
\end{figure}

If we want to put realistic numbers for the index $\nu$, we can use the constraint provided by the value of
the  spectral index $n_{s}$ of the CMB anisotropies (as summarized in Ref. \cite{2010arXiv1001.4538K}). It leads to $\nu\approx 40$. Such a value  invalidates the previous expansion.
On the left panel Fig. \ref{NuEff} we then compare the de Sitter case (lower curve) to the power law inflation case for $\nu$ varying from $35$ to $50$ (that corresponds to $n_{s}$ in the $0.94-0.96$ range). The amplitude of the vertex value, and therefore the coupling amplitude of test fields, is found to be amplified at the end of inflation by a factor that can be as large as 7 showing that quantitative results obtained in the de Sitter limit could be quite misleading!
% This comes as a surprising result, showing that although qualitatively robust (e.g. de Sitter is a genuine limit case of power law inflation when the index goes to infinity), quantitative results obtained from the de Sitter case are not very robust and can be changed by a huge factor.

\subsection{The case of a massive field}

We pursue the exploration of the dependence of the results on the background dynamics with a more realistic, and data favored, inflationary model, that is inflation produced by a massive field. 

In this paragraph we assume that the background evolution is derived from an inflaton potential $V_{\rm inf}(\varphi)$ such, 
\begin{equation}
V_{\rm inf}(\varphi)=\frac{1}{2} m^2 \varphi^2,
\end{equation}
where $\varphi$ is the inflaton field. In order to simplify the calculation and to obtain closed forms, we
assume that the slow roll solution can be used during the whole period of interest for our calculations. Simple numerical investigations show that this is indeed a good approximation. We then have,
\begin{equation}
\varphi(t)=\varphi_{0}-\sqrt{\frac{2}{3}}\,m\ \Mpl t
\end{equation}
where $t$ is the physical time, $\varphi_{0}$ is the value of $\varphi$ at horizon crossing for the scales of interest and $\Mpl$ is the Planck mass. The time evolution of $H$ can similarly be expressed as
\begin{equation}
H(t)=H_{0}-H_{1}^2 t
\end{equation}
with
\begin{equation}
H_{0}=\frac{1}{\sqrt{6}}\frac{m}{\Mpl}\varphi_{0},\ \ \hbox{and}\ \ H_{1}^2=\frac{m^2}{3}. 
\end{equation}
The initial value $\varphi_{0}$ is determined by the number of foldings one imposes on the inflationary period. Within those approximations, we have, $N_{e}={\varphi_{0}^2}/{4\Mpl^2}$ and the time at the end of inflation is given by $t_{\rm end}=2{N_{e}}/{H_{0}}$.

We then obtain,
\begin{eqnarray}
g_{0}^{\rm mf}(t,t'))&=&\frac{1}{m}\sqrt{\frac{\pi}{2}}\exp\left[\frac{9}{2}\frac{H_{0}^2}{m^2}+H_{0}(t+t')-\frac{H_{1}^2}{2}(t^2+t'^2)\right]\nonumber\\\
&&\times\left[\erfi \left(\frac{-H(t')}{\sqrt 2 m}\right)
-\erfi \left(\frac{-H(t)}{\sqrt 2 m}\right) \right]
\end{eqnarray}
where $\erfi$ is the imaginary error function and the vertex value is finally given by
\begin{equation}\label{nueffmassive}
\nue^{\rm mf}(N_{e})=2\lambda \frac{N_{e}^2}{H_{0}^2}\ _{2}F_{2}\left[1,1;\frac{3}{2},2;-3 N_{e}\right],
\end{equation}
where $\ _{2}F_{2}$ is the hypergeometric function. On the bottom panel of Fig. \ref{NuEff}  we show how the effective vertex value evolves with time for two different hypothesis for $N_{e}$. Once again, we see that the amplitude of the Non-Gaussian vertex value is large compared to the de Sitter case and amplified by a factor slightly smaller than 10.

What is then to be concluded for the $k_{i}$ dependence of the effective $f_{\rm NL}$ (or $\tau_{\rm NL}$) parameters? This explicit dependence cannot be derived from the resolution of the classical system but requires the full resolution of the quantum system. However the results obtained here suggest that the wave-mode dependence of the non-Gaussian parameters is suppressed by a factor of about 7 compared to the de Sitter case.  This implies that the shape dependence is at percent level whereas the global scale dependence is at sub-percent level.

\section{The global structure of the correlation functions at tree order}

These calculations can obviously be pursued in a perturbative expansion, in particular for the calculation of higher order correlation functions. Unsurprisingly one expects a close relation between the quantum calculation and the classical calculations. More precisely we expect the leading order in $g$ in the quantum calculations to be identical to the classical calculations. 

In the following we illustrate these calculations with the four-point function in case of a cubic potential. 

\subsection{Quantum calculation}

Let us start with the expression of the leading term of four-point correlation function in case of a cubic potential. It naturally involves two 
vertices and it is therefore given by the expression of the expectation value of $\mg \chi_{\vk_{1}}\dots \chi_{\vk_{4}}\md_{c}$
at second order in $H^{(I)}$. As an application of the relation (\ref{QobsExp}) it is thus given by,
\begin{eqnarray}
\mg \chi_{\vk_{1}}\dots \chi_{\vk_{4}}\md_{c}&=&-\frac{1}{a^4(\eta)}
\int_{\eta_{0}}^{\eta}\dd\eta'
\int_{\eta_{0}}^{\eta'}\dd\eta''\nonumber\\
&&\times\mg0\vert
\left[\left[\hu_{\vk_{1}}\dots\hu_{\vk_{4}},H^{(I)}(\eta')\right],H^{(I)}(\eta'')\right]
\vert 0\md
\end{eqnarray}
Replacing in this expression $H^{(I)}$ by its value 
%we have,
%\begin{eqnarray}
%\mg \chi_{\vk_{1}}\dots \chi_{\vk_{4}}\md_{c}&=&-\frac{1}{a^4(\eta)}\left(\frac{\lambda}{3!}\right)^2
%\nonumber\\
%&&\hspace{-3.5cm}\times\int\frac{\dd^3\vx'}{(2\pi)^3}\frac{\dd^3\vx''}{(2\pi)^3} 
%{\dd^3\vk'_{1}}\dots{\dd^3\vk'_{3}}
%{\dd^3\vk''_{1}}\dots{\dd^3\vk''_{3}}\nonumber\\
%&&\hspace{-3.5cm}\times\ e^{\ii\vx'.\sum_{i}\vk'_{i}+\ii\vx''.\sum_{i}\vk''_{i}}
%\int_{\eta_{0}}^{\eta}\dd\eta'
%\int_{\eta_{0}}^{\eta'}\dd\eta''a(\eta')a(\eta'')
%\nonumber\\
%&&\hspace{-3.5cm}
%\times\mg0\vert
%\left[\left[\hu_{\vk_{1}}(\eta)\dots\hu_{\vk_{4}}(\eta),\hu_{\vk'_{1}}(\eta')\dots\hu_{\vk'_{3}}(\eta')\right],\hu_{\vk''_{1}}(\eta'')\dots\hu_{\vk''_{3}}(\eta'')\right]
%\vert 0\md
%\end{eqnarray}
%The explicit computation of such a contribution 
leads to,
\begin{eqnarray}
\mg \chi_{\vk_{1}}\dots \chi_{\vk_{4}}\md_{c}&=&-\frac{{\lambda}^2(2\pi)^3}{a^4(\eta)}
\Dirac\left(\sum_{i=1} ^4\vk_{i}\right)\nonumber\\
&&\hspace{-3cm}\times
\int_{\eta_{0}}^{\eta}\dd\eta' a(\eta')
\int_{\eta_{0}}^{\eta'}\dd\eta'' a(\eta'')
\left[G_{k_{1}}(\eta,\eta')G_{k_{2}}(\eta,\eta')-\hbox{c.c.}\right]\nonumber\\
&&\hspace{-3cm}\times
\left[G_{k_{3}}(\eta,\eta'')G_{k_{4}}(\eta,\eta'')G_{\vert\vk_{1}+\vk_{2}\vert}(\eta',\eta'')-\hbox{c.c.}\right]+\sym
\end{eqnarray}
where $\hbox{c.c.}$ stands for complexe conjugate and where $\sym$ means that all symmetric terms should be included (there are $6$
such contributions in total). When the $G$ functions are replaced by their expression in terms of their real and imaginary parts, the result scales either in $P^3\,g^2$ or in $P\,g^4$. We naturally expect the first type of terms to dominate for super-horizon scales. These are those that we consider in the following. That leads to the superhorizon expression of the correlation function,
\begin{eqnarray}
\mg \chi_{\vk_{1}}\dots \chi_{\vk_{4}}\md_{c}^{\rm SH}&=&\frac{{\lambda}^2(2\pi)^3}{a^4(\eta)}
\Dirac\left(\sum_{i=1} ^4\vk_{i}\right)\nonumber\\
&&\hspace{-3.cm}\times
\int_{\eta_{0}}^{\eta}\dd\eta' a(\eta')
\int_{\eta_{0}}^{\eta'}\dd\eta'' a(\eta'')
\left[P_{k_{1}}(\eta,\eta')g_{k_{2}}(\eta,\eta')+g_{k_{1}}(\eta,\eta')P_{k_{2}}\right]\nonumber\\
&&\hspace{-3.cm}\times
\left[P_{k_{3}}(\eta,\eta')g_{k_{4}}(\eta,\eta'')P_{\vert\vk_{1}+\vk_{2}\vert}(\eta',\eta'')+\sym\right.\nonumber\\
&&\hspace{-2.cm}\left.+
P_{k_{3}}(\eta,\eta'')P_{k_{4}}(\eta,\eta'')g_{\vert\vk_{1}+\vk_{2}\vert}(\eta',\eta'')+\sym\right].
\label{QuantumDec}
\end{eqnarray}
In this expression the $\sym$ terms account respectively for 11 and 5 other terms. As can be seen there are two types of contributions. Each one corresponds to a different term - and graphs - in the classical picture. This is what we show explicitly in the following.

\subsection{Classical calculation}

\begin{figure}[htbp]
\begin{center}
\epsfig{file=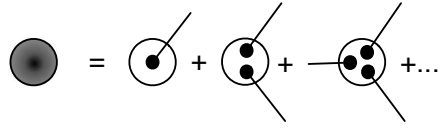,scale=.7}
\caption{Formal diagrammatic representation of the expansion (\ref{chiExp}). Each term has a specific order in the initial free field expression. This order is represented by the number of dots in the circle. The Wick theorem imposes that dots should be associated by pairs. Each dot has value $\chi^{(i)}_{k_{i}}$, each circle represents the operator, $\nu_{p}/p!\int\dd^3\vk\Dirac(\vk-\sum\vk_{i})$ where the explicit value of each vertex depends on the background dynamics as explained in the text.}
\label{ChiExp}
\end{center}
\end{figure}

\begin{figure}[htbp]
\begin{center}
\epsfig{file=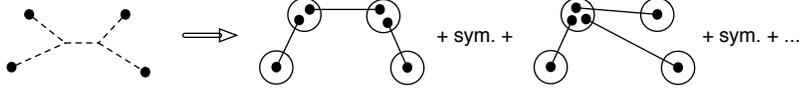,scale=.5}
\caption{Diagrammatic representation of the four-point correlation function. The left side represents the standard quantum Feynmann diagram (note that in case of an expanding background, this representation is a oversimplified since there are different propagators to consider). The right hand side represents its super-Hubble limit where such a contribution is split into 2 parts. Those diagrams are obtained by gluing different terms of the expansion depicted in Fig. \ref{ChiExp} following the Wick theorem (dots are associated by pairs). Once the integration over $\vk_{i}$ is done, each line then carries a power spectrum factor $P_{k_{i}}(\eta)$.}
\label{Corr4Diag}
\end{center}
\end{figure}

The calculation of $\mg \chi_{\vk_{1}}\dots \chi_{\vk_{4}}\md_{c}$ in the classical picture is indeed given by a further expansion of
$\chi$ in terms of the \textsl{initial field} value, e.g.
\begin{equation}
\chi_{\vk}=\chi^{(i)}_{\vk}+\chi^{(ii)}_{\vk}+\chi^{(iii)}_{\vk}+\dots\label{chiExp}
\end{equation}
where $\chi^{(i)}$ is linear in the initial field, $\chi^{(ii)}$ is quadratic. The expansion is therefore different from  the expansion of Eq. (\ref{uexpansion}).
As a result the expression of $\mg \chi_{\vk_{1}}\dots \chi_{\vk_{4}}\md_{c}$  at lowest order in $\lambda$ will be
\begin{eqnarray}
\mg \chi_{\vk_{1}}\dots \chi_{\vk_{4}}\md_{c}=
\mg \chi^{(i)}_{\vk_{1}}\chi^{(ii)}_{\vk_{2}}\chi^{(ii)}_{\vk_{3}}\chi^{(i)}_{\vk_{4}}\md_{c}+
\mg \chi^{(i)}_{\vk_{1}}\chi^{(iii)}_{\vk_{2}}\chi^{(i)}_{\vk_{3}}\chi^{(i)}_{\vk_{4}}\md_{c}+\sym.\label{ClassicDec}
\end{eqnarray}
where $\sym$ accounts for respectively 5 and 3 other terms. The expressions of $\chi^{(p)}$ can be expressed recursively in terms of the classical Green function and in terms of the free fields $\chi^{(i)}$. More precisely we have,
\begin{equation}
\chi^{(ii)}_{\vk}(\eta)=-\frac{\lambda}{2(2\pi)^3}\int^{\eta}{\dd\eta'}{\dd^3\vk'}\,g_{k}(\eta,\eta')\,a(\eta')\,u^{(0)}_{\vk'}(\eta')u^{(0)}_{\vk-\vk'}(\eta')
\end{equation}
and
\begin{eqnarray}
\chi^{(iii)}_{\vk}(\eta)
&=&\frac{\lambda^2}{2(2\pi)^3}
\int^{\eta}{\dd\eta'}{\dd^3\vk'}\,g_{k}(\eta,\eta')\,a(\eta')\,u_{\vk-\vk'}^{(i)}(\eta')\,u_{\vk'}^{(ii)}(\eta')\\
&=&\frac{\lambda^2}{2(2\pi)^6}
\int^{\eta}{\dd\eta'}{\dd^3\vk'}\,g_{k}(\eta,\eta')\,a(\eta')\,u_{\vk-\vk'}^{(i)}(\eta')\nonumber\\
&&\hspace{-1cm}\times\int^{\eta'}{\dd\eta''}\frac{\dd^3\vk''}{(2\pi)^{3/2}}\,g_{k'}(\eta',\eta'')a(\eta'')\,
u^{(i)}_{\vk'-\vk''}(\eta'')u^{(i)}_{\vk'-\vk''}(\eta'')
\end{eqnarray}
It is straightforward to see that the two contributions in (\ref{ClassicDec}) precisely correspond to the 2 terms in (\ref{QuantumDec}). 

In the super-horizon regime, it is furthermore natural to take the super-horizon limit behavior for those quantities. That implies that
\begin{eqnarray}
\mg \chi_{\vk_{1}}\dots \chi_{\vk_{4}}\md_{c}^{\rm SH}&=&\frac{{\lambda}^2}{a^8(\eta)}(2\pi)^3
\Dirac\left(\sum_{i=1} ^4\vk_{i}\right)\nonumber\\
&&
\left[P_{k_{1}}P_{k_{3}}P_{\vert\vk_{1}+\vk_{2}\vert}\ 
\left(\int_{\eta_{0}}^{\eta}\dd\eta'
g_{0}(\eta,\eta')a^3(\eta')\right)^2+\sym\right.
\nonumber\\
&&\hspace{-4cm}\left.+
P_{k_{1}}P_{k_{2}}P_{k_{3}}\ a(\eta)\int_{\eta_{0}}^{\eta}\dd\eta' g_{0}(\eta,\eta')a^2(\eta')\int_{\eta_{0}}^{\eta'}\dd\eta'' g_{0}(\eta',\eta'')a^3(\eta'')+\sym
\right]
\end{eqnarray}
where the power spectrum are to be taken at equal time $\eta$. Again one should properly take into account the symmetry factors.
This expression further simplifies in,
\begin{eqnarray}
\mg \chi_{\vk_{1}}\dots \chi_{\vk_{4}}\md_{c}^{\rm SH}&=&\frac{\nu^2_{\rm eff.}}{a^6}\,P_{k_{1}}P_{k_{3}}P_{\vert\vk_{1}+\vk_{2}\vert}
+\sym\nonumber\\
&&+\frac{\nu_{3}}{a^6}\,P_{k_{1}}P_{k_{2}}P_{k_{3}}+\sym
\end{eqnarray}
where $\nu_{\rm eff.}$ is given by Eq. (\ref{nueexp}) and where $\nu_{3}$ is,
\begin{equation}
\nu_{3}=\frac{\lambda}{a(\eta)}\int^{\eta}\dd\eta'\,g_{0}(\eta,\eta')a(\eta')^3\,\nu_{\rm eff.}(\eta').\label{cum4tree}
\end{equation}
The classical picture naturally leads to the construction of a whole set of such vertices that can be built recursively.
This is what is shown below.

\subsection{The tree structure}

The general structure of the correlation functions of the classical problem taken at leading order  is well known. Similar structures have been encountered in other aspect of cosmological instability growth (see Refs. \cite{1984ApJ...279..499F,1992ApJ...392....1B,2002PhR...367....1B}). It corresponds to a tree structure where each vertex point corresponds to a given order of $\chi$ in the expansion of Eq. (\ref{chiExp}). To be more precise,
\begin{equation}
\mg\chi_{\vk_{1}}\dots\chi_{\vk_{n}}\md_{c}=\sum_{\{p_{i}\},\sum_{i}p_{i}=2(n-1))}\mg\chi^{(p_{1})}_{\vk_{1}}\dots\chi^{(p_{n})}_{\vk_{n}}\md_{c}\label{cumpformal}
\end{equation}
where the sum is to be done for all values of $\{p_{i}\}$ that are integers and with the constraints that $\sum_{p_{i}}=2(n-1)$. This constraint comes from the fact that in order to connect $n$ points, one needs at least $n-1$ links and therefore the order of the 
product is at east $2(n-1)$ whereas terms with more than $n-1$ links are subdominant and are therefore not considered here.
When represented in a diagrammatic expansion, as in Fig. \ref{Corr4Diag}, all contributing terms of Eq. (\ref{cumpformal}) then form a tree. 

For the problem of interest here the calculation further simplifies since each order of $\chi$ in the super-Hubble limit is
given by a simple vertex value multiplied by a convolution of the zeroth order terms,
\begin{eqnarray}
\chi^{(p)}_{\vk}(\eta)&=&\frac{\nu_{p}(\eta)}{p!(2\pi)^{3(p-1)}}\int\Dirac\left({\vk-\sum_{i}\vk_{i}}\right)\dd^3\vk_{1}\dots\dd^3\vk_{p}\nonumber\\
&&\times \chi^{(1)}_{\vk_{1}}(\eta)\dots\chi^{(1)}_{\vk_{p}}(\eta).
\end{eqnarray}
where $\nu_{2}=\nue$. 
In other words the super-Hubble classical properties of the fields will be determined, at tree order, by a discrete set of numbers: the vertices that appear in all those constructions. This result extends the form given in Eq. (\ref{cum4tree}) for the four-point function. All high order correlation functions form then a genuine tree structure characterized by the expression of the power spectrum and the vertex values.

And because the vertices are all independent of $k$, it is actually relatively easy to 
derive them at once. 
Consider for that the related system defined by the motion equation, $u_{0}=a(\eta)\chi_{0}(\eta)$,
\begin{equation}
u_{0}''(\eta)-\frac{a''(\eta)}{a(\eta)}\,u_{0}(\eta)=-a^3(\eta)\frac{\dd V}{\dd \chi}[u_{0}(\eta)/a(\eta)].\label{uk0evol}
\end{equation}
which obviously exhibits the very same late time growing mode and same Green function\footnote{At this stage we obtain what a direct application of the $\delta N$ formalism would have given but with a very specific interpretation and as a mean to provide a controlled approximation of the general expression of the fourth cumulant given in Eq. (\ref{fourthcumgeneral}) and of \textsl{tree-order} correlators in general.}. We thus have,
\begin{equation}
\chi_{0}^{(p)}(\eta)=\frac{\nu_{p}(\eta)}{p!}\left[\chi_{0}^{(1)}(\eta)\right]^p.
\end{equation}
so that the solution of this equation, taken at the time of interest, is nothing but the vertex generating function\footnote{These generating functions can be used to build local PDFs (see \cite{2002PhR...367....1B} for details on this construction in general and \cite{2002PhRvD..66j3506B} for this peculiar case) 
but it is beyond the scope of this paper to present such a construction.}. This identification was already pointed out in \cite{2002PhRvD..66j3506B} although with a looser connexion to the quantum field calculations. Solving (\ref{uk0evol}) might prove difficult in general. This equation can however be solved numerically, or even analytically in some peculiar cases.  It is particularly interesting to rewrite the motion equation (\ref{uk0evol})  in time variable $t$ and field variable $\chi_{0}$, such that
\begin{equation}
\ddot\chi_{0}+3H\dot\chi_{0}=-\frac{\dd V}{\dd \chi}\left[\chi_{0}\right].\label{chi0evol}
\end{equation}
The two equations are clearly equivalent. More precisely we define $\mG(t,\chi_{0})$ as the solution of (\ref{chi0evol}) with the initial conditions $\chi_{0}(t_{0}=0,\tau)=\tau$ and $\dot \chi_{0}(t_{0}=0,\tau)=0$ which corresponds to set $\mG$ to the growing mode of the evolution equation. The generating function $\mG(\tau)$ of the vertices is then nothing but
\begin{equation}
\mG(\tau)\equiv \sum_{p}\frac{\nu_{p}(\eta)}{p!}\left[\chi_{0}^{(1)}(\eta)\right]^p=\chi_{0}(t_{\rm end},\tau).
\end{equation}
where $t_{\rm end}$ is the time at the end of inflation.

\begin{figure}[htbp]
\begin{center}
\epsfig{file=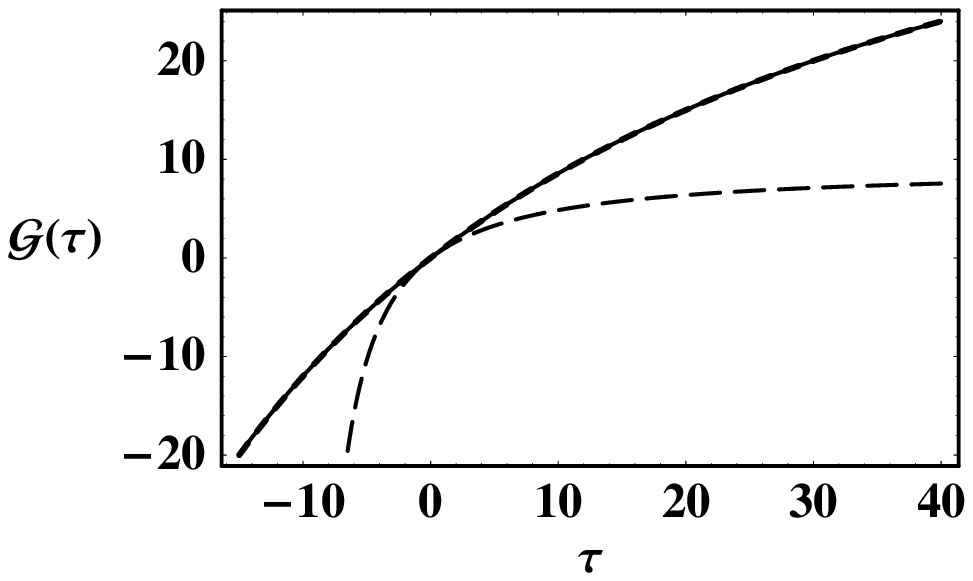,width=7cm}
\epsfig{file=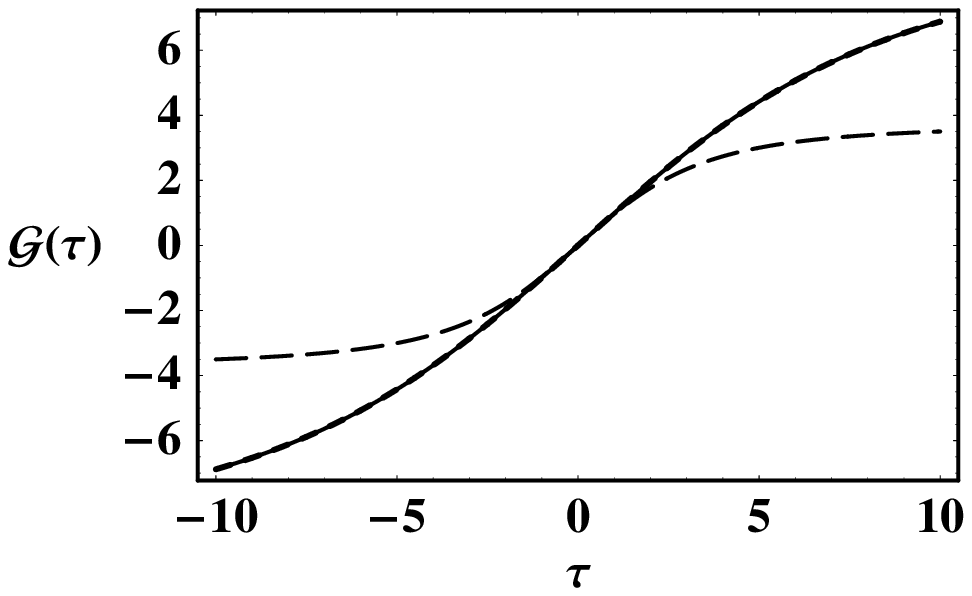,width=7cm}
\caption{The shape of the vertex generating function $\mG(\tau)$ for a de Sitter background (solid lines) compared to the approximate form of Eq. (\ref{mGdSexp}), dotted line, for a cubic potential (top) and quartic potential (bottom). The results are also compared
to the power law inflation case (dashed lines). The curves correspond to $\lambda N_{e}=0.1$ and $N_{e}=60.$ The power law exponent corresponds to a spectral index of $n_{s}=0.95$}
\label{GenV3}
\end{center}
\end{figure}

We present in Figs. \ref{GenV3} the resulting shapes of the generating functions $\mG(\tau)$ for different cases. They are compared to approximate forms we derive below.

The obtention of  approximate forms is based on the observation that as long as the interaction term is small, the first term of the left hand side of (\ref{chi0evol}) is negligible compared to the second (e.g. the time evolution if $\chi_{0}$ is weak). Within this approximation the evolution equation of $\chi_{0}$ is simply
\begin{equation}
3 H\dot\chi_{0}=\frac{\dd V}{\dd \chi}\left[\chi_{0}\right].\label{chi0evol2}
\end{equation}
which can be formally solved in
\begin{equation}
\int_{\tau}^{G}\frac{\dd \chi_{0}}{V'(\chi_{0})}=-\int_{t_{0}}^{t_{\rm end}}\,\frac{\dd t}{3 H}.
\end{equation}
For a potential in $V(\chi)=\lambda \chi^{p}/p!$ we are left with
\begin{equation}
\mG(\tau)=\tau\left[1+\lambda\tau^{p-2}\frac{p-2}{(p-1)!}\int_{t_{0}}^{t_{\rm end}}\frac{\dd t}{3 H}\right]^{\frac{1}{2-p}}.
\end{equation}
For a de Sitter background we then have
\begin{equation}
\mG^{\rm dS}(\tau)=\tau\left[1+\frac{p-2}{(p-1)!} \frac{\lambda\ N_{e}}{3 H^2}\,\tau^{p-2}\right]^{\frac{1}{2-p}}
\label{mGdSexp}
\end{equation}
after $t_{\rm end}-t_{0}$ has been replaced by $N_{e}/H$. For a power law inflation it leads to
\begin{equation}
\mG^{\rm pl.}(\tau)=\tau\left[1+\frac{p-2}{(p-1)!} \frac{\lambda\ \left[\exp(2N_{e}/\nu)-1\right]}{6 \nu H^2}\,\tau^{p-2}\right]^{\frac{1}{2-p}}.
\label{mGPLexp}
\end{equation}

Interestingly, for this type of potential, the structures of the $\tau$ dependence are  the same and independent on the background dynamics. They differ only by the value of the lowest order vertex, $\nue$. It means for instance that for a cubic potential $\nu_{3}$ scales like $\nu_{2}^2$, e.g. it grows like $N_{e}^2$ is the de Sitter limit. In other words, to a good approximation, the entire background dependence of the non-linear structure is encoded in the effective vertex expression that in the dependence given by the r.h.s of Eq. (\ref{nueexpression}).

\subsection{Conditions for large isocurvature fluctuations revisited}

In paragraphe \ref{Conditionsonlambda} conditions were given on $\lambda$ for isocurvature perturbations to exist
at horizon crossing. The result of the previous paragraph now suggests some validity rules for the perturbation theory calculations  
to be valid during the super-Hubble evolution. Indeed the previous calculations make sense only if 
of the nonlinear transform $\mG(\tau)$ is still close to a linear transform when its argument $\tau$ is of the order
of its typical value at horizon crossing, that is $H$. 

For a de Sitter background it implies that the dimensionless quantity that appears in the expression of $\mG^{\rm dS}(\tau)$, e.g.
${\lambda\ N_{e}}H^{p-4}/3$, is less than unity. It simply states that $\nu_{3}^{\rm dS}H^2$ should remain small.
The condition is now more stringent than the one derived in paragraphe \ref{Conditionsonlambda}. 
For other backgrounds we can then exploit 
the fact the functions $\mG$ are very similar once expressed in terms of $\nu_{3} \tau^2$. This suggests that
the validity conditions for other backgrounds simply reads $\nu_{3}H^2 \lesssim 1$ which introduces an extra factor 7 or so.
The conditions is slightly more stringent than for de Sitter case, it now reads $\lambda \lesssim 1/(7N_{e})$ for a quartic potential,
but leave significant room for the $\chi$ field to develop large fluctuations and yet develop significant non-Gaussianities.

\section{Conclusions and comments}

We have explored the dependence of the evolution of the high order correlation functions of a test scalar field with minimal coupling on the background dynamics. The exact resolution of this problem requires involved quantum field theory calculation and there are only a limited number of cases where the calculations can be fully completed. They correspond to a de Sitter background and to some specific cases of power law inflation. Some of these results were already known but a novel and effective method is presented here that allows to perform those calculations efficiently. The explicit $k$-dependence of the bispectrum and trispectrum  is derived for such cases for respectively a cubic or quartic potential. 

It is furthermore explicitly shown how the super-horizon evolution is equivalent to the mode coupling evolution of a classical system. It is then possible to derive the super-horizon evolution of the amplitude of the higher order spectra for arbitrary background. The result is encapsulated in Eq. (\ref{nueexpression}). This expression is computed for different classes of models such as general power law inflation or background of chaotic inflation. It is found that when the background evolution departs from a de Sitter behavior the coupling amplitude is amplified. 
For realistic cases, we found  that this amplification is about a factor 7. Those results suggest that the evolution is actually minimal when the background is de Sitter.

It is to be noticed however that the resolution of the related classical system does not permit, even within a perturbation scheme, to compute the exact $k_{i}$ dependence of the effective $\tau_{\rm NL}$ and $f_{\rm NL}$ parameters of the iso-curvature perturbations. Only in specific cases, where the quantum evolution can be computed, are those results known. If the background is close enough to de Sitter we found that the shape variations can induce variation of $\tau_{\rm NL}$ or $f_{\rm NL}$ parameters up to 7 percent when the number of efolds is about 50 from horizon crossing to the end of inflation. The scale dependence (e.g. dependence with respect to scale for a given configuration) is precisely given by $1/N_{e}$ where $N_{e}$ is the number of efolds. Results obtained for a different background suggest that these dependence are about 7 times smaller for the inflationary backgrounds investigated in this paper. Those dependences are expected to be transferred to those of the metric fluctuations for
certain models of multiple-hybrid inflation. Note however that for models where the transfer of modes takes place sooner, that is after a much smaller number of efolds, the $k$-dependence can be made arbitrarily larger. Its precise value depends however on the details of the model and needs a full quantum calculation to be derived.

We finally show that when one restricts the computation to tree order, once super-Hubble scales have been reached, the high-order correlation functions follow a genuine tree structure (see text for details) with $k$-independent vertices. Correlation functions
are then entirely encoded in the amplitude of the two-point correlation function and in the generating function of the vertices that can be explicitly computed for any backgrounds. Examples strongly suggest that, to a very good approximation,
the background effects are entirely encoded in the expression of the lowest order vertex so that higher order correlation functions, when estimated for super-horizon limits, can indeed be built from this vertex value only.

\ifcqg
\section*{References}
\else
\fi
\ifcqg
\bibliographystyle{h-physrev}
\else
\bibliographystyle{apsrev}
\fi

\bibliography{EarlyUniverse}

\end{document}